\documentclass{article}
\usepackage{dcase2019,amsmath,graphicx,url,times,booktabs, tabularx}
\pdfoutput=1

\def\relus{{\ensuremath{\operatorname{\mathop{ReLU6}\,}}}}
\def\conv{{\ensuremath{\operatorname{\mathop{conv2d}\,}}}}
\def\depthwise{{\ensuremath{\operatorname{\mathop{dwise}\,}}}}

\title{URBAN SOUND TAGGING USING CONVOLUTIONAL NEURAL NETWORKS}

\name{Sainath Adapa}
\address{FindHotel\\
    Amsterdam, Netherlands \\
     adapasainath@gmail.com}

\begin{document}

\ninept
\maketitle

\begin{sloppy}

\begin{abstract}
In this paper, we propose a framework for environmental sound classification in a low-data context (less than 100 labeled examples per class). We show that using pre-trained image classification models along with the usage of data augmentation techniques results in higher performance over alternative approaches. We applied this system to the task of Urban Sound Tagging, part of the DCASE 2019. The objective was to label different sources of noise from raw audio data. A modified form of MobileNetV2, a convolutional neural network (CNN) model was trained to classify both coarse and fine tags jointly. The proposed model uses log-scaled Mel-spectrogram as the representation format for the audio data. Mixup, Random erasing, scaling, and shifting are used as data augmentation techniques. A second model that uses scaled labels was built to account for human errors in the annotations. The proposed model achieved the first rank on the leaderboard with Micro-AUPRC values of 0.751 and 0.860 on fine and coarse tags, respectively.
\end{abstract}


\section{Introduction}
\label{sec:intro}

The IEEE AASP challenge on Detection and Classification of Acoustic Scenes and Events (DCASE) \footnote{http://dcase.community/}, now in its fifth edition, is a recurring set of challenges aimed at developing computational scene and event analysis methods. In Task 5, Urban Sound Tagging, the objective is to predict the presence or absence of 23 different tags in audio recordings. Each of these tags represents a source of noise and thus a cause of noise complaints in New York City. Solutions for this task, such as the one proposed in this paper, will help inspire the development of solutions for monitoring, analysis, and mitigation of urban noise.

\section{Related work}
\label{sec:related}

The current task of Urban Sound Tagging is part of the broader research area of Environmental Sound Classification \cite{houix2012lexical}. Convolutional neural networks (CNNs) that use Log-scaled Mel-spectrogram as the feature representation have been proven to be useful for this use case \cite{salamon2017deep, Lasseck2018}, and have also achieved leading performance in recent DCASE tasks \cite{Jeong2018, Akiyama2019, Chen2019}. Extensions to the CNN framework, in the form of Convolutional Recurrent Neural Networks (CRNNs) have been proposed \cite{zhang2019attention}. Transformation of the raw audio waveform into the Mel-spectrogram representation is a "lossy" operation \cite{wyse2017audio}. As such, there has been ongoing research into evaluating alternatives such as using Scattering transform \cite{salamon2015feature},  Gammatone filter bank \cite{zhang2019attention} representations, as well as directly employing one-dimensional CNN on the raw audio signal \cite{abdoli2019end}. Operating in the context of noisy labels \cite{fonseca2019learning} or in a low-data regime \cite{pons2018training} (both of which are properties of the present task) are two other active research areas in this domain. One particular approach for dealing with small labeled datasets is the usage of pre-trained models to generate embeddings that can be used for downstream audio classification tasks. VGGish\cite{hershey2017cnn}, SoundNet\cite{aytar2016soundnet}, and L\textsuperscript{3}-Net\cite{cramer2019look} are examples of such models.

\section{Dataset}
\label{sec:dataset}

For this challenge, SONYC \cite{Bello2019sonyc} has provided 2351 recordings as part of the \textit{train set}, and 443 recordings as a part of the \textit{validate set}. All the recordings, acquired from different acoustic sensors in New York City, are Mono channel, sampled at 44.1kHz, and are ten seconds in length. The private \textit{evaluation set} consisted of 274 recordings. Labels for these recordings were revealed only at the end of the challenge. A single recording might contain multiple noise sources. Hence, this is a task of multi-label classification.

The 23 noise tags, termed \textit{fine-grained tags}, are further grouped into a list of 7 \textit{coarse-grained tags}. This hierarchical relationship is illustrated in Figure \ref{fig:taxonomy}. Each recording was annotated by three Zooniverse\footnote{https://www.zooniverse.org/} volunteers. Additional annotations, specifically for \textit{validate set}, were performed by the SONYC team members and ground truth is then agreed upon by the SONYC team. Since the \textit{fine-grained tags} are not always easily distinguishable, annotators were given the choice of assigning seven tags of the form "other/unknown" for such cases. Each of these seven tags termed "incomplete tags," correspond to a different coarse category.

\begin{figure}[h]
\centering
\includegraphics[scale=0.45]{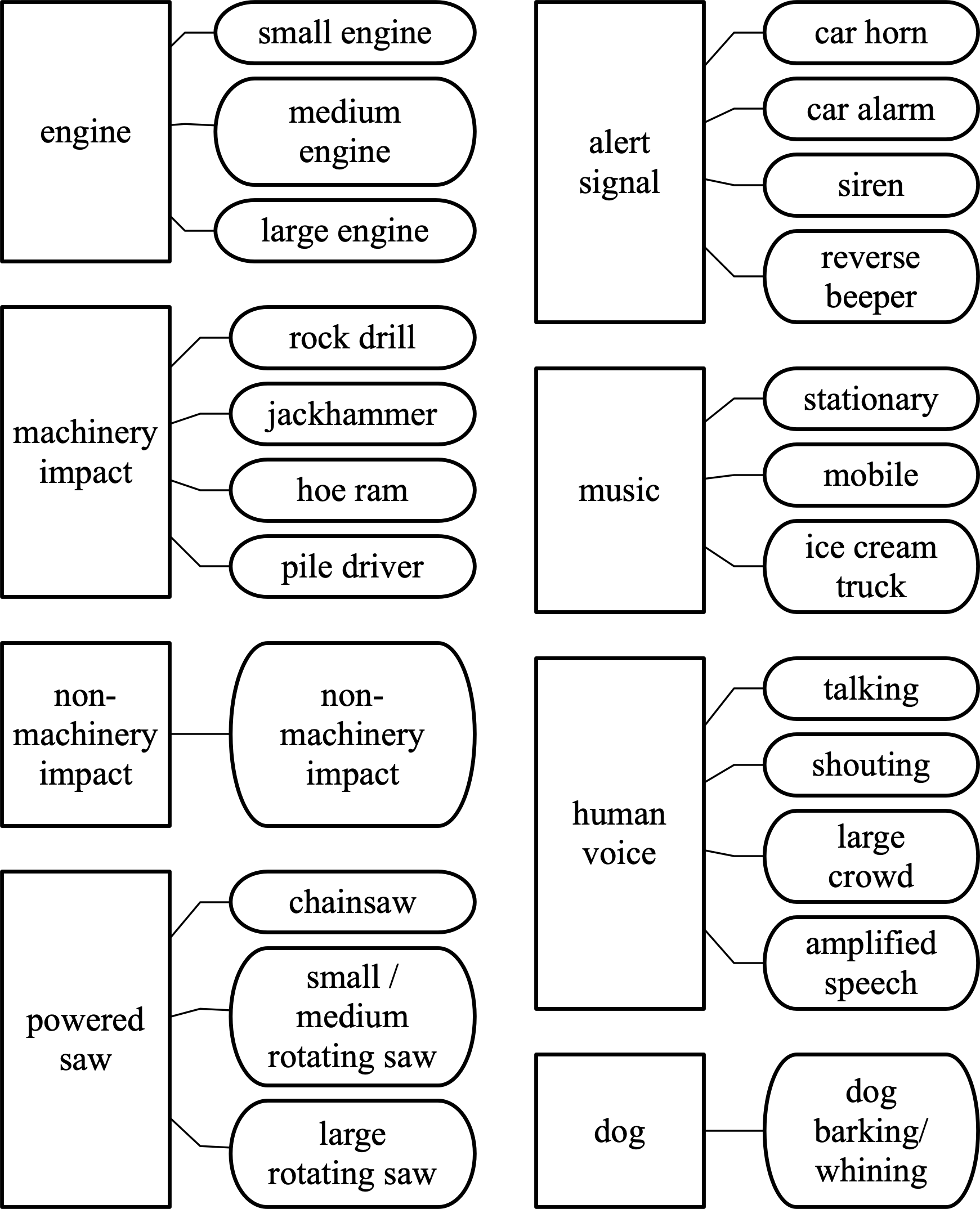}
\caption{Hierarchical taxonomy of tags. Rectangular and round boxes respectively denote coarse and fine tags respectively. \protect\cite{dcase2019task5}}
\label{fig:taxonomy}
\end{figure}

\section{Proposed framework}
\label{sec:framework}

\subsection{CNN Architecture}
\label{ssec:architecture}

In this work, we use a modified form of MobileNetV2 \cite{sandler2018mobilenetv2}. The architecture of MobileNetV2 contains a 2D convolution layer at the beginning, followed by 19 \textit{Bottleneck residual blocks} (described in Table \ref{fig:bottlenec_block_table}). Spatial average of the output from the final residual block is computed and used for classification via a Linear layer.

The proposed model makes few modifications to the above-described architecture. The input Log Mel-spectrogram data is sent to the MobileNetV2 after passing it through two convolution layers. This process transforms the single-channel input into a three-channel tensor to match the input size of original MobileNetV2 architecture. Instead of the spatial average, Max pooling is applied to the output from the final residual block. Additionally, the single linear layer at the end is replaced by two linear layers. The full architecture is described in Table \ref{mobilenet:arch}.

\begin{table}
\centering
    \begin{tabular}{c|c|c}
    Input & Operator & Output\\
    \toprule
    $h \times w \times k$ & 1x1 \conv, \relus & $h \times w \times tk$\\
    $h \times w \times tk$& 3x3 \depthwise s=$s$, \relus & $\frac{h}{s} \times \frac{w}{s} \times (tk)$\\ 
    $\frac{h}{s} \times \frac{w}{s} \times tk$ & linear 1x1 \conv & $\frac{h}{s} \times \frac{w}{s} \times k'$\\
    \toprule
    \end{tabular}
    \caption{{\em Bottleneck residual block} transforming from $k$ to $k'$ channels, with stride $s$, and expansion factor $t$. \protect\cite{sandler2018mobilenetv2}}
\label{fig:bottlenec_block_table}
\end{table}

\begin{table}[]
\centering
\vspace{0pt}
\begin{tabular}{c|c|c|c|c}
Operator & $t$ & $c$ & $n$ & $s$ \\ \hline
\textbf{conv2d} & \textbf{-} & \textbf{10} & \textbf{1} & \textbf{1} \\
\textbf{conv2d} & \textbf{-} & \textbf{3} & \textbf{1} & \textbf{1} \\
conv2d & - & 32 & 1 & 2 \\
bottleneck & 1 & 16 & 1 & 1 \\
bottleneck & 6 & 24 & 2 & 2 \\
bottleneck & 6 & 32 & 3 & 2 \\
bottleneck & 6 & 64 & 4 & 2 \\
bottleneck & 6 & 96 & 3 & 1 \\
bottleneck & 6 & 160 & 3 & 2 \\
bottleneck & 6 & 320 & 1 & 1 \\
conv2d 1x1 & - & 1280 & 1 & 1 \\
\textbf{maxpool} & \textbf{-} & \textbf{1280} & \textbf{1} & \textbf{-} \\
\textbf{linear} & \textbf{-} & \textbf{512} & \textbf{1} & \textbf{-} \\
\textbf{linear} & \textbf{-} & \textbf{k} & \textbf{1} & \textbf{-} \\ \hline
\end{tabular}
\caption {Each line describes a sequence of 1 or more identical (modulo stride)  layers, repeated $n$ times. All layers in the same sequence have the same number $c$ of output channels. The first layer of each sequence has a stride $s$ and all others use stride $1$. All spatial convolutions use $3\times 3$ kernels (except for the first two which use $1\times 1$ kernels). The expansion factor $t$ is always applied to the input size as described in Table~\ref{fig:bottlenec_block_table}. Modifications to the MobileNetV2 architecture are highlighted in bold.}
\label{mobilenet:arch}
\end{table}

\subsection{Initialization with Pre-trained weights}
\label{ssec:imagenet}
In many fields, including in the acoustic area, CNNs exhibit better performance with an increase in the number of layers \cite{eigen2013understanding, seide2011conversational}. However, it has been observed that deeper neural networks are harder to train and prone to overfitting, especially in the context of limited data \cite{szegedy2015going}.

Many of the \textit{fine-grained tags} have less than 100 training examples with positive annotations, thus placing the current task into a \textit{low-data regime} context \cite{pons2018training}. Since the proposed architecture has a large  (24) number of layers, we initialized all the unmodified layers of the network with weights from the MobileNetV2 model trained on ImageNet \cite{russakovsky2015imagenet, mv2weights}. Kaiming initialization \cite{he2015delving} is used for the remaining layers. Since the domain of audio classification is different from image classification, we do not employ a Fine-tuning approach \cite{girshick2014rich} here. All the layers are jointly trained from the beginning. When all the layers with ImageNet weights were frozen at that parameters, the model performed worse than the baseline model (Section \ref{sec:results}) showing the need for joint training of the whole network.

The rationale behind the use of ImageNet weights is that the kind of filters that the ImageNet based model has learned are applicable in the current scenario of Spectrograms as well. Especially the filters in the initial layers that detect general patterns like edges and textures\cite{zeiler2014visualizing} are easily transferable to the present case. With the described initialization, we noticed faster and better convergence (illustrated in Figure \ref{fig:imagenet_kaimin} and Table \ref{table:imagenet_kaimin_final_loss}) when compared to initializing all the layers with Kaimin initialization. Similar gains were observed previously in the context of Acoustic Bird Detection \cite{Lasseck2018}.

Other pre-trained models such as ResNeXt\cite{xie2017aggregated}, and EfficientNet\cite{tan2019efficientnet} were also tested. The observed metrics were at the same level as the MobileNetV2 architecture. Since the performance is similar, MobileNetV2 was chosen as it has the least number of parameters among the models tried.

\begin{figure}[!htbp]
\centering
\includegraphics{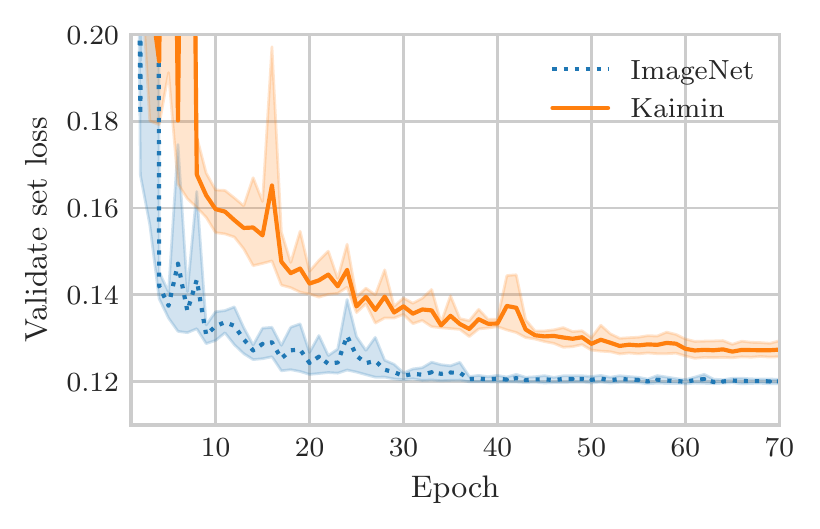}
\caption{Trajectory of \textit{validate set} loss during training, demonstrating that using pre-trained ImageNet weights results in faster convergence.}
\label{fig:imagenet_kaimin}
\end{figure}

\begin{table}[]
\begin{tabular}{lll}
\toprule
 & \multicolumn{1}{c}{\begin{tabular}[c]{@{}c@{}}ImageNet\\ pre-trained\\ weights\end{tabular}} & \multicolumn{1}{c}{\begin{tabular}[c]{@{}c@{}}Kaimin\\ initialization\end{tabular}} \\ \toprule
\textit{train set} loss & 0.1401 $\pm$ 0.0017 & 0.1493 $\pm$ 0.0019 \\
\textit{validate set} loss & 0.1200 $\pm$ 0.0008 & 0.1266 $\pm$ 0.0022 \\
\toprule
\end{tabular}
\caption{Final Binary Cross-entropy loss values at the end of training. 5 repetitions of training runs from scratch were performed.}
\label{table:imagenet_kaimin_final_loss}
\end{table}

\subsection{Preprocessing and Data augmentation}
\label{ssec:preprocessing}

The proposed model uses Log Mel-spectrogram as the representation format for the input data. Librosa \cite{brian_mcfee_2019_2564164} toolbox was used to compute the Mel-spectrogram. For the Short-time Fourier transform (STFT), \textit{window length} of 2560 and \textit{hop length} of 694 was used. For the Mel-frequency bins computation, the lowest and the highest frequencies were set at 20Hz and 22050Hz, respectively, with the number of bins being 128.\footnote{https://www.kaggle.com/daisukelab/fat2019\_prep\_mels1} No re-sampling or additional preprocessing steps were performed.

Several data augmentation techniques were used to supplement the training data. Deformations such as Time stretching and Pitch shifting that were previously shown to help in sound classification were employed \cite{salamon2017deep}. Also, image augmentation methods such as Random rotate, Grid distortion \cite{2018arXiv180906839B}, and Random erasing \cite{zhong2017random} were used. Mixup \cite{zhang2017mixup}, an approach that linearly mixes two random training examples was used as well. Table \ref{table:dataaug} shows the impact of Data augmentations, when each of the methods were applied separately.

\begin{table}
\centering
  \vspace{6pt}
  \centering
\begin{tabular}{lll}
\toprule
 & \begin{tabular}[c]{@{}l@{}}Fine-level\\ Micro-AUPRC\end{tabular} & \begin{tabular}[c]{@{}l@{}}Coarse-level\\ Micro-AUPRC\end{tabular} \\
\toprule
No data augmentation & 0.716 & 0.819 \\
Only Mixup & 0.745 & 0.840 \\
Only Random erasing & 0.732 & 0.820 \\
Only Random rotate & 0.728 & 0.832 \\
Only Shifting time & 0.719 & 0.822 \\
Only Grid distortion & 0.753 & 0.842 \\
\begin{tabular}[c]{@{}l@{}}Pitch shifting and\\ Time stretching\end{tabular} & 0.732 & 0.834 \\
All the techniques & 0.772 & 0.855 \\
\toprule
\end{tabular}
\caption{Performance on the \textit{validate set}, demonstrating the gains due to data augmentation}
\label{table:dataaug}
\end{table}

\subsection{Re-labeling}
\label{ssec:relabeling}

For the \textit{validate set}, we have access to both the ground truth and the three sets of annotations by Zooniverse volunteers. When the ground truth of a label is positive, 36\% of annotations (by Zooniverse volunteers) do not match with the ground truth. If the quality of the labels can be improved, it is quite possible that the accuracy of the model can be increased as well. Hence, a logistic regression model that takes the annotations as input and estimates the ground truth label was developed. This model was trained on the \textit{validate} set, and then the ground truth estimate for the \textit{train set} was generated. Table \ref{fig:relabeleg} shows a sample of predictions from the model.

\begin{table}[]
\centering
\begin{tabular}{cccc}
\hline
\begin{tabular}[c]{@{}c@{}}Coarse\\ label\end{tabular} & \begin{tabular}[c]{@{}c@{}}Fine\\ label\end{tabular} & \begin{tabular}[c]{@{}c@{}}Positive\\ annotations\\ count\end{tabular} & \begin{tabular}[c]{@{}c@{}}Predicted\\ score\end{tabular} \\ \hline
music & uncertain & 1 & 0.10 \\
music & uncertain & 3 & 0.98 \\
music & stationary & 2 & 0.88 \\
powered saw & chainsaw & 3 & 0.98 \\
machinery impact & - & 0 & 0.05 \\ \hline
\end{tabular}
\caption{Predictions for few cases from the automatic re-labeling model}
\label{fig:relabeleg}
\end{table}

\section{Model Training}

\subsection{Evaluation metric}
Area under the precision-recall curve using the micro-averaged precision and recall values (Micro-AUPRC) is used as the classification metric for this task. Micro-F1 and Macro-AUPRC values are reported as secondary metrics. Detailed information about the evaluation process is available on the task website \cite{dcase2019task5}.

\subsection{Training}

Two models were trained for this challenge:

\textbf{M1:} The first model generates probabilities for both the fine and coarse labels. During training, whenever the annotation is "unknown/other", loss for the fine tags corresponding to this coarse tag was masked out. Hence, this model does not generate predictions for \textit{uncertain} fine labels. Since there are three sets of annotations for each training example, one by each Zooniverse volunteer, the loss is computed against each annotation set separately. Average of the three loss values is taken as the final loss value for a training example.

\textbf{M2:} For the second model, predictions from the re-labeling model described in Section \ref{ssec:relabeling} are used as labels. This model generates probabilities for both the fine and coarse labels, including the \textit{uncertain} fine labels.

Both the models use identical input data representation and employ the same data augmentation techniques (mentioned in Section \ref{ssec:preprocessing}). They also use Binary Cross-entropy loss as the optimization metric. The models are trained on the \textit{train set} using the \textit{validate set} to determine the stopping point.

Training was done on PyTorch \cite{paszke2017automatic}. AMSGrad variant of the Adam algorithm \cite{kingma2014adam, reddi2019convergence} with a learning rate of 1e-3 was utilized for optimization. Whenever the loss on \textit{validate} set stopped improving for five \textit{epochs}, the learning rate was reduced by a factor of 10. Regularization in the form of Early stopping was used to prevent overfitting \cite{prechelt1998early}. At the time of prediction, test-time augmentation (TTA) in the form of Time shifting was used.

\section{Results}
\label{sec:results}

The baseline system mentioned on the task page \cite{dcase2019task5} computes VGGish embeddings \cite{hershey2017cnn} of the audio files and builds a multi-label logistic regression model on top of the embeddings. For this baseline system, a label for an audio recording is considered positive if at least one annotator has labeled the audio clip with that tag. Table \ref{table:task5} shows the performance of the baseline system compared against the proposed models on the private \textit{evaluation set}. The proposed models\footnote{https://github.com/sainathadapa/urban-sound-tagging} exhibit improved Micro-AUPRC values for both \textit{fine-grained} and \textit{coarse-grained} labels when compared against the baseline model. Moreover, it can be observed that re-labeling didn't prove effective; it helped improve the Micro-F1 score significantly, but it didn't help raise Micro-AUPRC or Macro-AUPRC.

Class-wise AUPRC performance is reported in Table \ref{table:classwiseperf}. The modified MobileNetV2 architecture improves over the Baseline model performance for all classes (except one) at both coarse and fine-level prediction. In the case of coarse-level prediction, the AUPRC performance for "Powered saw" is lesser than that of Baseline.

\begin{table}
\centering
  \vspace{6pt}
  \centering
  \resizebox{\columnwidth}{!}{%
  \begin{tabular}{l p{0.8cm}p{0.8cm}p{0.8cm}p{0.8cm}p{0.8cm}p{0.8cm}}
    \toprule
    & \multicolumn{3}{c}{\textbf{\textsc{Fine-level}}} & \multicolumn{3}{c}{\textbf{\textsc{Coarse-level}}} \\
    & \multicolumn{3}{c}{\textbf{\textsc{prediction}}} & \multicolumn{3}{c}{\textbf{\textsc{prediction}}} \\
	\cmidrule(lr){2-4} \cmidrule(lr){5-7} 
	& \small{Macro AUPRC} & \small{Micro F1} & \small{Micro AUPRC} & \small{Macro AUPRC} & \small{Micro F1} & \small{Micro AUPRC} \\
	\midrule
 \begin{tabular}[c]{@{}l@{}}Baseline\end{tabular} & 0.531 & 0.450 & 0.619 & 0.619 & 0.664 & 0.742 \\
 \begin{tabular}[c]{@{}l@{}}M1\end{tabular} & 0.645 & 0.484 & 0.751 & 0.718 & 0.631 & 0.860 \\
 \begin{tabular}[c]{@{}l@{}}M2\end{tabular} & 0.622 & 0.575 & 0.721 & 0.723 & 0.745 & 0.847 \\
	\bottomrule
\end{tabular}}
\caption{Performance on the private \textit{evaluation set}}
\label{table:task5}
\end{table}

\begin{table}
\centering
  \vspace{6pt}
  \centering
  \resizebox{\columnwidth}{!}{%
  \begin{tabular}{l p{0.8cm}p{0.8cm}p{0.8cm}p{0.8cm}p{0.8cm}p{0.8cm}}
    \toprule
    & \multicolumn{3}{c}{\textbf{\textsc{Coarse-level}}} & \multicolumn{3}{c}{\textbf{\textsc{Fine-level}}} \\
    & \multicolumn{3}{c}{\textbf{\textsc{prediction}}} & \multicolumn{3}{c}{\textbf{\textsc{prediction}}} \\
	\cmidrule(lr){2-4} \cmidrule(lr){5-7} 
	& \small{Baseline} & \small{M1} & \small{M2} & \small{Baseline} & \small{M1} & \small{M2} \\
	\midrule
 \begin{tabular}[c]{@{}l@{}}Engine\end{tabular} & 0.832 & 0.888 & 0.878 & 0.638 & 0.665 & 0.673 \\
 \begin{tabular}[c]{@{}l@{}}Machinery impact\end{tabular} & 0.454 & 0.627 & 0.578 & 0.539 & 0.718 & 0.604 \\
 \begin{tabular}[c]{@{}l@{}}Non-machinery impact\end{tabular} & 0.170 & 0.361 & 0.344 & 0.182 & 0.362 & 0.374 \\
 \begin{tabular}[c]{@{}l@{}}Powered saw\end{tabular} & 0.709 & 0.684 & 0.643 & 0.478 & 0.486 & 0.378 \\
 \begin{tabular}[c]{@{}l@{}}Alert signal\end{tabular} & 0.727 & 0.897 & 0.875 & 0.543 & 0.858 & 0.832 \\
 \begin{tabular}[c]{@{}l@{}}Music\end{tabular} & 0.246 & 0.404 & 0.586 & 0.168 & 0.289 & 0.351 \\
\begin{tabular}[c]{@{}l@{}}Human voice\end{tabular} & 0.886 & 0.947 & 0.949 & 0.777 & 0.841 & 0.833 \\
\begin{tabular}[c]{@{}l@{}}Dog\end{tabular} & 0.929 & 0.937 & 0.931 & 0.922 & 0.936 & 0.931 \\
\bottomrule
\end{tabular}}
\caption{Class-wise AUPRC on the private \textit{evaluation set}}
\label{table:classwiseperf}
\end{table}

\section{Conclusions and Future directions}
\label{sec:conclusions}

In this paper, we presented our solution to Task 5 (Urban Sound Tagging) of the DCASE 2019 challenge. Our approach involved using a pre-trained image classification model and modifying it for audio classification. We also employed data augmentation techniques to help with the training process. This resulted in our model achieving Micro-AUPRC values of 0.751 and 0.860 on Fine and Coarse tags, respectively thus obtaining the first rank on the leaderboard. We thus demonstrated that impressive gains could be made when compared to using audio embeddings, even in a low-resource scenario such as the one presented here.

As noted in \cite{Gousseau2019}, AUPRC only partially correlates with cross-entropy, i.e., decrease in Binary cross-entropy loss may not always result in increase in AUPRC. Exploring loss functions that are more related to AUPRC metric is an avenue for improvement. Depending on the type of class to be predicted, different input representations (such as STFT, HPSS, Log-Mel) might be better \cite{Bai2019}. Thus, an ensemble model that uses these different representations can surpass the one proposed in this paper. This ensemble can also involve models that use VGGish or L\textsuperscript{3}-Net embeddings.

\bibliographystyle{IEEEtran}

%
%
%
%
%
%
%
%
%

\end{sloppy}
\end{document}